\documentclass[a4paper, 12pt]{article}
\usepackage[utf8]{inputenc}
\usepackage{titletoc}
\usepackage{indentfirst}
\usepackage{setspace}
\usepackage{lipsum}
\usepackage{fancyhdr}
\usepackage[left=2.5cm, right=2cm, top=1in, bottom=1in, bindingoffset=0cm]{geometry}
\usepackage{titlesec}
\usepackage{natbib}
\usepackage{etoolbox}
\usepackage{graphicx}
\usepackage{amsfonts}
\usepackage{array}
\usepackage{multirow}
\usepackage{amsmath}
\usepackage{subfig}
\usepackage{systeme}
\graphicspath{ {./images/} }
\titleformat*{\section}{\normalfont\fontsize{14}{19}\selectfont}
\renewcommand{\contentsname}{\centering Contents}
\begin{document}

\pagestyle{fancy}
\fancyhf{}
\lhead{FX MARKET VOLATILITY}
\fancyfoot[R]{\thepage}
\begin{spacing}{1.5}

\begin{titlepage} 

	\centering 
	
	\vspace*{\baselineskip} 
	
	
	\rule{\textwidth}{1.6pt}\vspace*{-\baselineskip}\vspace*{2pt} 
	\rule{\textwidth}{0.4pt} 
	
	\vspace{0.5\baselineskip} 
	
	{\Large FX Market Volatility} 
	
	\vspace{0.5\baselineskip} 
	
	\rule{\textwidth}{0.4pt}\vspace*{-\baselineskip}\vspace{3.2pt} 
	\rule{\textwidth}{1.6pt} 
	
	\vspace{1\baselineskip} 
	
	
	\large Senior Thesis
	
	\vspace*{2\baselineskip} 
	
	
    By
	
	\vspace{0.5\baselineskip} 
	
	{\scshape\Large Anton Koshelev \\ Group 165 \\} 
	
	\vspace{0.5\baselineskip} 
	
	National Research University Higher School of Economics \\
	School of World Economy and International Affairs \\
	
	\vfill 
	
	{\large Academic advisor: \\ }
	{\scshape\Large Prof. Vasily Solodkov \\}
	
	\vspace*{6\baselineskip} 
	Moscow, 2020
	
\end{titlepage}
\addtocounter{page}{1}
\renewcommand*\contentsname{\centering{Contents}}
\renewcommand{\listtablename}{\centering{List of Tables}}
\renewcommand{\listfigurename}{\centering{List of Figures}}
\dottedcontents{section}[1em]{\bfseries}{2.9em}{1pc}
\tableofcontents
\newpage
\listoffigures
\listoftables
\newpage

\setlength{\parindent}{0.5in}
\addcontentsline{toc}{section}{Abstract}
\section*{\centering{\textbf{Abstract}}}
This paper aims at solving FX market volatility modeling problem and finding the most becoming approach to this task. Validity of two competing approaches, classical econometric generalized conditional  heteroscedasticity and mathematical (singular spectrum analysis and dynamical systems stability analysis) are tested on major currency pairs (EUR/USD, USD/JPY, GBP/USD) and unique high-frequency USD/RUB data. The study shows that both  mathematical tools, understudied in econometric discourse, have a great potential in scope of discussed problematic, as for all experiments covered in this research, both of them show promising results.\\

\textit{Keywords:} volatility modeling, foreign exchange market\\
\hfill \break
\hfill \break

\addcontentsline{toc}{section}{Introduction}
\section*{\centering{\textbf{Introduction}}}

\begin{figure}
    \centering
    \includegraphics[scale = 0.25]{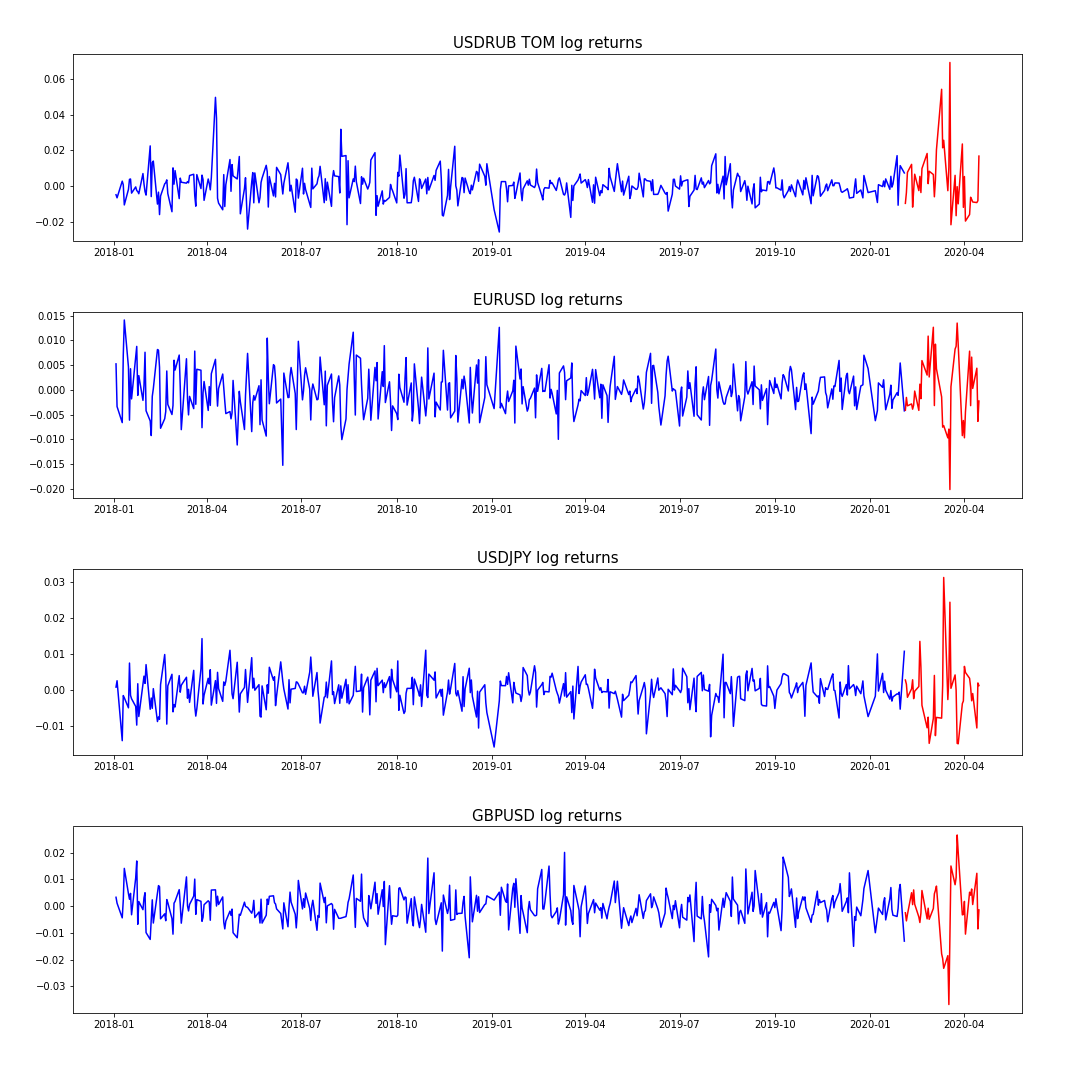}
    \caption{FX market volatility in March-April 2020}
    \label{log_returns_general}
\end{figure}

Foreign exchange (FX) market is one of the major financial markets in contemporary word, with average daily volume of approximately 6.6 trillion US dollars \citep{wooldridge2019fx}, which exceeds any other segment of the global financial system. On over the counter (OTC) and centralized parts of FX markets floating currency rates are determined by existing demand and supply. These rates are widely and constantly used by banks, non-bank financial organizations, companies involved in export or import and households, which, in turn, constitute the whole economy. Thus, abrupt changes in FX rates, which became frequent with the crash of Bretton-Woods system \citep{klein2012exchange}, pose a great threat to the stability of the world economy, which makes the nature of FX market volatility worth investigating and studying. Indeed, as it is empirically shown by \citet{klein2012exchange} reserve currency pairs have been exhibiting higher volatility since the introduction of free-floating regimes in 1974 and their statement find evidence in the newest FX market statistics of March 2020, when severe volatility took place in nearly all segments of foreign exchange market (figure \ref{log_returns_general}).\\

In this paper, we will focus on the concept of observed volatility and determine it as standard deviation of asset’s returns over a specified period of time, as it is done in the broader literature \citep{andersen2009realized}, however, methods employed in this paper strongly deviate from the mainstream econometric thought. The topic of financial market volatility has been the subject of econometric investigation for a long period of time: \citet{andersen2009handbook} provide us with the conventional definitions of volatility types and \citep{bollerslev1994arch} summarize classical econometric techniques of volatility clustering and forecasting. Additionally, \citet{bergomi2015stochastic} discusses modern advances in stochastic volatility modelling, which is another method of understanding main drivers of assets’ volatility. On the other hand, prominent mathematical base has been developed in the second half of the XX century, aiming at finding determinants of complex non-linear systems (without any connection to volatility modeling task), which were successfully used in various spheres of natural sciences \citep{malinetskii2000modern}. This apparatus was employed later in \citet{hassani2010predicting}, \citet{bohm1999expectations} and a few other papers for tasks of quantitative finance, such as realized volatility forecasting. Despite high potential of iconoclastic mathematical concepts and promising results of early works, this approach is still underexamined in scholar discourse of econometricians, who prefer focusing on well-known autoregressive conditional heteroskedasticity (ARCH) models, which place severe constraints on the data used \citep{poon2005practical}. Therefore, it seems useful to shift the focus to non-linear mathematical models, which do not have such questionable assumptions and have already proved their efficiency in other scientific fields.\\

The primary goal of this study is to cover the methodological gap in the modern econometric research of FX market volatility by means of applying abstract mathematical framework to financial time series, describing the state of FX market. In the first part of this paper, a proper definition and classification of volatility types is presented and  a detailed overview of classical econometric methods of volatility modeling is given. After that, unconventional mathematical tools suitable for this task are discussed. Second part of this research presents empirical comparison of discussed approaches. Two types of “new” models are built: singular spectrum analysis model (SSA) and a model of a dynamical system, which are compared to the most widespread GARCH model using historical data for major currency pairs and unique high frequency time series for USD/RUB provided by Moscow Exchange. Finally, we summarize results, which prove the main hypothesis of mathematical models’ supremacy in comparison with popular econometrics to be trustworthy, when two sets of models are compared in terms of predictive ability.\\

\newpage
\addcontentsline{toc}{section}{1. Literature Review}
\section*{\centering{\textbf{1. Literature Review}}}
\addcontentsline{toc}{subsection}{1.1 Stylized Facts about Volatility}
\subsection*{\centering{\textbf{1.1 Stylized Facts about Volatility}}}

Let us start the research of volatility by defining this process. In a broader sense, volatility is associated with asset’s price variability. Precisely, econometric literature denotes volatility as asset returns' standard deviation over time interval $T$. As changes in the returns play a significant role in assets’ pricing, risk management and portfolio allocation tasks, volatility modeling has been dominating econometric literature for the past 40 years, since the Nobel prize winning article of \citet{engle1982autoregressive} describing autoregressive conditional heteroscedasticity (ARCH) model came out. During the next four decades, a number of adjusted and alternative techniques were developed. ARCH cluster of models was extended to cover external (mainly macroeconomic) factors \citep{engle2007good} and blended with autoregressive moving average (ARMA) process \citep{degiannakis2008arfimax}, while scholars also started extracting market expectations of future volatility from current option prices \citep{andersen2009handbook}, calling it ‘implied volatility’. What is more, stochastic volatility theory emerged, which does not directly rely on pre-determined past volatility values \citep{engle2007good}. These and other methods will be reviewed in more detail later in this paper.\\

Additionally, a number of stylized facts about financial market volatility behavior were documented by econometricians and economists. Firstly, it was found out that volatility demonstrates some persistence, which means that periods of high or low return deviation tend to last for some time and do not vanish immediately \citep{engle2007good}, \citep{baillie1996fractionally}. Secondly, volatility is widely considered to be a mean-reverting process, indicating the asymptotic historical level of volatility, that ‘attracts’ temporary deviations \citep{engle2007good}. Thirdly, it is still a matter of a heated debate between scholars, whether there is an asymmetric impact of return innovations of different signs on volatility level. While \citet{black1976studies}, \citet{engle1993measuring} prove it to be true for equities, no evidence of asymmetry is found for foreign exchange market. Furthermore, it was shown by \citet{andersen1998deutsche} that external factors may significantly influence volatility on the case of DM/USD currency rate and US macroeconomic data releases and \citet{engle2007good} came to the same conclusion with Dow Jones stock market index and short-term interest rate changes. Next, it was empirically proved by \citep{engle2007good} that unconditional distribution of an asset’s historic volatilities has too heavy tails (with kurtosis far greater than 3.0) to be named Gaussian. These main points help us to start drawing a proper picture of a studied problematic.\\

Now, let us move on to the widely-used methods of volatility modeling and forecasting, as well as on their benefits and drawbacks.\\

\addcontentsline{toc}{subsection}{1.2 Mainstream Volatility Models}
\subsection*{\centering{\textbf{1.2 Mainstream Volatility Models}}}

An instrument prevailing in econometric discourse as well as in the basic economics is autoregressive conditional heteroskedasticity (ARCH) models and their derivatives. This group of models assumes that conditional variance $h_t$ (equation \ref{ARCH}) is a function of squared random variable occurrences taken for the specific time window of length $q$:

\begin{equation} \label{ARCH}
    \ h_t = \alpha_0 + \sum\limits_{j=1}^q \alpha_j \varepsilon^{2}_{t-j}
\end{equation}
where $y_t$ is an observable random variable, $\varepsilon_t = y_t - \mathbb{E}(y_t) = z_t h^{1/2}_t, a_0 > 0, a_j \geq 0,  j = 1, ..., 1-q,$ and $a_q > 0 $.\\

Nowadays, generalized ARCH (GARCH) process is used the most in econometric and financial literature. It implies extended specification of conditional variance $h_t$, which is now also depends linearly from both $\varepsilon$ and its own lags of orders 1 to $p$ (equation \ref{GARCH}).

\begin{equation} \label{GARCH}
    \ h_t = \alpha_0 + \sum\limits_{j=1}^q \alpha_j \varepsilon^{2}_{t-j} + \sum\limits_{j=1}^p \beta_j h_{t-j}
\end{equation}

Since its introduction in 1986 by \citet{bollerslev1986generalized} a number of GARCH upgrades have been developed: non-linear GARCH models like smooth transaction GARCH and threshold GARCH, time-varying GARCH, Markov-switching GARCH, integrated GARCH, exponential and multivariate models of generalized conditional heteroskedasticity, which are covered in \citet{terasvirta2009introduction}. However, according to the research conducted by \citet{terasvirta2009introduction}, the most popular model of GARCH is a simple GARCH(1,1), which only looks at the first lags of $\varepsilon$ and $h$. Coefficients of regressors $\alpha_i$ and $\beta_i$ can be estimated using maximum likelihood method and sufficient condition for the GARCH process to be weakly stationary is $\sum \alpha_i + \sum \beta_i < 1$, as it is stated in \citet{terasvirta2009introduction}.\\

(G)ARCH models have both strong sides and weak points. Although GARCH has proved to be robust for short-term conditional volatility modelling, it assumes a symmetrical effect of both positive and negative innovations in time series, which does not align with empirical observations as it is written in \citet{engle2007good}. These authors also show (G)ARCH dependency on data points frequency in terms of the model specification when the same asset is studied, but time steps vary. What is more, \citet{engle2007good} focus their attention on the fact that even if a number of assets’ conditional volatilities tend to be described by GARCH model precisely, portfolios constituted of the same assets are not necessarily described by this model properly. Finally, the idea of volatility persistence described earlier is violated by GARCH unless $p$ parameter is large enough.\\

All these shortcomings of classic ARCH models pushed the vector of econometric thought to the direction of stochastic volatility (SV), which is an alternative way of volatility drivers identification, also brightly presented in modern financial econometrics. As it is stated in \citet{jungbacker2009parameter}, SV is a model with observation equation (\ref{SV_1}) and state equation (\ref{SV_2}):

\begin{equation} \label{SV_1}
    y_t = \mu + exp(\frac{1}{2}h_t)\varepsilon_t, \varepsilon_t \sim NID(0, 1)
\end{equation}

\begin{equation} \label{SV_2}
    h_{t+1} = \gamma + \phi h_t + \eta_t, \eta_t \sim NID(0, \sigma^2_\eta)
\end{equation}

The estimation of next period volatility is dependent on equation coefficients $\gamma$, $\phi$, $\eta_t$, which can be found with the help of maximum likelihood function. However, \citet{jungbacker2009parameter} argue that for the class of SV models, this task cannot be solved analytically, which motivates an econometrician for the use of approximations, numerical methods and simulations (like Monte Carlo method). Although SV approach first discussed by \citet{taylor2008modelling}, \citet{harvey1994multivariate} is widely thought to be more accurate than ARCH-type models \citep{koopman2005forecasting}, they have a series of significant drawbacks. Firstly, the estimation of parameters depends heavily on the used algorithm, which can be computationally intense. Secondly, SV models still have assumptions regarding the nature of variables’ distributions \citep{satchell2011forecasting}. Finally, \citet{jungbacker2009parameter} draw reader’s attention to the fact that different methods for SV specification are not applicable in all situations.\\

Financial industry also knows another way of FX volatility modeling. As current market information might be to some extent represented in current prices, option quotes may give us an insight regarding underlying asset’s future volatility. This concept is called ‘implied volatility’. Precise discussion of this technique is outside the scope of this paper, but reader can refer to \citet{christensen1998implied}, where the topic is fully covered. \citet{engle2007good} condemn this type of market-expectation-based forecast for its reliance on a specific model of option pricing, which probably incorporates time-varying volatility risk premium and, hence, is biased.\\

\addcontentsline{toc}{subsection}{1.3 Alternative Ways of Volatility Modeling}
\subsection*{\centering{\textbf{1.3 Alternative Ways of Volatility Modeling}}}

In the second half of XX century, parallel to ARCH and SV models, different methods in applied mathematics and statistics were developed, which were not initially designed to address the problems of financial time series analysis. Among them one can find singular spectrum analysis and dynamical systems. Singular spectrum analysis (SSA) is a set of statistical tools aimed at time series analysis by means of matrix decomposition. It is completely uncorrelated with previously reviewed ARMA and ARCH methods, wavelet or Fourier transform, which, at the first glance, face the same task. SSA is a model free methodology and it rests on few assumptions regarding the time series used as an input. The main prerequisite is that initial vector is a sum of a signal and noise \citep{golyandina2013singular}. We will study this assumption in a closer manner later in this chapter.\\

SSA, which is a part of a signal processing family consisting of principal component analysis (PCA), projection pursuit, independent component analysis (ICA) and others, has a variety of applications. It can be used for smoothing (also done by Fourier transform), noise reduction, extraction of trends of different resolution, periodicity recognition and volatility estimation \citep{golyandina2013singular}.\\

According to \citet{golyandina2001analysis}, SSA algorithm consists of several steps: matrix decomposition and trend reconstruction. Assume that we observe a vector (time series) $\mathbb{X}_N = (x_1, x_2, ..., x_N)$ which is decomposed into a trajectory matrix $X$ by the usage of $L$ lagged vectors $X_i = (x_i, ..., x_{i+L-1})^{T}$ $(i = 1, 2, ..., K$ where $K = N-L+1)$ as its columns. Then, matrix $XX^T$ becomes a subject of eigendecomposition (it other words, singular value decomposition). By doing this, one can acquire $L$ eigenvalues and $L$ eigenvectors of $XX^T$, from which a sub-sample of size $r$ is taken. Thus, $L$-dimensional data ${X_1, X_2, ..., X_K}$ is transferred into a subspace of a lower dimension $(r < L)$. After that, Hankel $\tilde{X}$ matrix is received after diagonal averaging process, and time series $(\tilde{x}_1, \tilde{x}_2, ..., \tilde{x}_N,)$, which is taken from $\tilde{X}$ is seen as an approximation of $\mathbb{X}_N$.\\

The only one significant assumption of SSA approach is the idea of data separability. \citet{golyandina2013singular} state that this suggests that it is possible to split the singular value decomposition (SVD) of the trajectory matrix $X$ into two sets so that the sum of terms within the sets gives trajectory matrices $X_1$ and $X_2$ of the series $\mathbb{X}_1$ and $\mathbb{X}_2$, which, in sum, give initial time series $\mathbb{X}$. As SVD is not unique, different types of time series separability arise. On the one hand, weak separability assumes that there exists at least one SVD, which satisfies the definition of separability. On the other hand, strong separability assumes that this holds for any singular value decomposition of an initial trajectory matrix. In practice, it can be difficult to prove separability of time series. To solve this problem, \citet{golyandina2013singular} deploy independent component analysis (ICA) technique before SSA, as it gives an opportunity to find independent signals in initial time series. ICA concept and algorithms are fully covered by \citet{hyvarinen1999survey} and one can study SSA theory in detail in \citet{golyandina2001analysis} and \citet{golyandina2013singular}.\\

Another non-financial approach to volatility modeling is named dynamical system analysis. This is a mathematical framework in the sphere of non-linear dynamics and chaos, which was soundly developed and improved in the second half of XX century by various schools. Our approach can be summarized as reviewing realized volatility time series as a reflection of some unknown complex and unobservable system, which has a set of properties and features that we can study by the analysis of an observable volatility time series. Moreover, these unknown properties of a dynamical system are able to give us an insight into the future behavior of its projection.\\

As it is stated in \citet{malinetskii2000modern}, each dynamical system can be viewed as an autonomous system of differential equations with two main parts: phase space $\mathbb{P}$ and a group of its transformations $\varphi^t(x)$, where $t$ denotes time, which can be assumed discrete or continuous, and $x$ is a starting point. Frequently, $\mathbb{P}$ is thought to be a n-dimensional Euclidean space or n-dimensional torus. Additionally, it is very important, that for any point (vector) $x \in \mathbb{P}$ there is only one unique point (vector) $\varphi^t(x) \in \mathbb{P}$. What is more, a key concept that we will study is a path of a dynamical system passing through point $x$, which equals to a manifold $\{\varphi^t(x)\}$, where $t$ takes all possible values. There are various possible paths of a system, from the simplest fixed points with $\varphi^t(x_0) = x_0$ for all $t$ and periodic paths $\varphi^{t+T}(x) = \varphi^t(x)$ to the most complicated ones. It is also worth stating that paths of a dynamical system do not intersect.\\

Dynamical system's paths can be stable or unstable in terms of their reaction to the small external exposure on the starting parameters of the system. Indeed, as it is done in \citet{malinetskii2000modern}, this small exposure can be seen as a kind of noise, which is an intrinsic part of reality. If this noise is unable to change the final outcome dramatically, the dynamical system acquires scientific value and becomes more attractive in terms of experiment reproducibility. Again, this stability (reproducibility) can be discussed from various viewpoints. Firstly, one can talk about a single path and its reaction to small changes in the starting point, which corresponds to the ordinary Lyapunov stability. Secondly, stability is sometimes seen as stable asymptotic behavior of an ensemble of paths. In other words, all trajectories originating in some area of $\mathbb{P}$ will be pulled to some invariant set. Simultaneously, ordinary Lyapunov stability of each path from the ensemble is not assumed (Lyapunov invariant set stability). Finally, there are other types of stabilities (discussed in \citet{strogatz2001nonlinear}), which are outside the scope of this paper.\\  

In this study we will use Lyapunov exponents derived from observable time series with the help of an algorithm presented in \citet{rosenstein1993practical} to determine dynamical system's stability or instability on a moving window, which size is also to be determined. With the knowledge of trajectory nature we will try to model future price perturbations for different time horizons in foreign exchange market.\\

\newpage
\addcontentsline{toc}{section}{2. Empirical Results}
\section*{\centering{\textbf{2. Empirical Results}}}

\addcontentsline{toc}{subsection}{2.1 Hypothesis, Metrics and Data}
\subsection*{\centering{\textbf{2.1 Hypothesis, Metrics and Data}}}

Above mentioned methods of FX volatility modeling (SSA and dynamical systems) are unconventional and understudied in financial econometrics literature, although they have already proved to be effective in other scientific fields, specifically in signal processing in case of SSA and in modeling of biological systems, physical processes and expansion of infectious diseases in field of non-linear dynamics. At the same time, classical econometric approach, heavily exploited for the last 40 years, has a number of disadvantages, strong assumptions and limitations, which were discussed in the previous chapter. This brings us to the main hypothesis $\mathbb{H}_0$ that in the task of volatility modeling SSA and dynamical system approaches would demonstrate better results in comparison with well-known autoregressive conditional heteroskedasticity models. Still, the problematic seems equivocal, as we need to establish what we understand by volatility and what is the metric of a model's suitability for the task. \\

In this paper, we define volatility as the standard deviation of hourly logarithmic returns $r_i = ln(\frac{P_i}{P_{i-1}}), r = \{r_0, r_1, ..., r_N\}$ within a trading day $i$ (equation \ref{volatility}):\\

\begin{equation} \label{volatility}
    \sigma_i = \sqrt{\frac{1}{T-1} \sum\limits_{t=1}^T (r_t- \overline{r})^{2}}
\end{equation}

As $\sigma_i$ is a real variable, the goodness of fit can be measured with the use of such metrics as mean squared error (MSE), mean absolute error (MAE) and others. However, due to specifics of Lyapunov exponent methodology, when the overall stability of the dynamical system's path is determined, we will shift our task in the classification domain with class $I$ representing an upper trend of daily volatility and class $II$ denoting an opposite return development, i.e. downside volatility trend. Real value output of GARCH model can easily be transformed into class by sign function $sign(\widehat{\sigma_i} - \sigma_{i-1})$, where $\widehat{\sigma_i}$ is a predicted value. \\ 

\begin{figure}
    \centering
    \includegraphics[scale = 0.75]{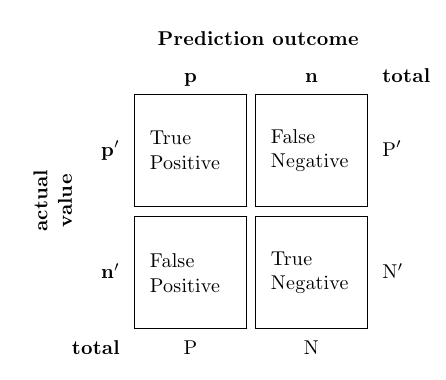}
    \caption{Confusion matrix}
    \label{confusion_matrix}
\end{figure}

In case of classification, the number of possible metrics is extensive \citep{scikit-learn}, but the majority is based on a confusion matrix \citep{fawcett2006introduction} displayed in figure \ref{confusion_matrix} (class $I$ is written as $p$ and class $II$ as $n$).\\

We will use the following classifier metrics: accuracy (equation \ref{accuracy}) and F1 score (eq. \ref{f1}) based on precision (eq. \ref{precision}) and recall (eq.\ref{recall}). They will help to evaluate different aspects of models, see them from different angles and make final statement regarding $\mathbb{H}_0$ hypothesis.\\

\begin{equation} \label{accuracy}
    accuracy = \frac{TP + TN}{TP + TN + FP + FN}
\end{equation}

\begin{equation} \label{precision}
    precision = \frac{TP}{TP + FP}
\end{equation}

\begin{equation} \label{recall}
    recall = \frac{TP}{TP + FN}
\end{equation}

\begin{equation} \label{f1}
    f1 = 2 * \frac{Precision * Recall}{Precision + Recall}
\end{equation}

\begin{figure}
    \centering
    \includegraphics[scale = 0.30]{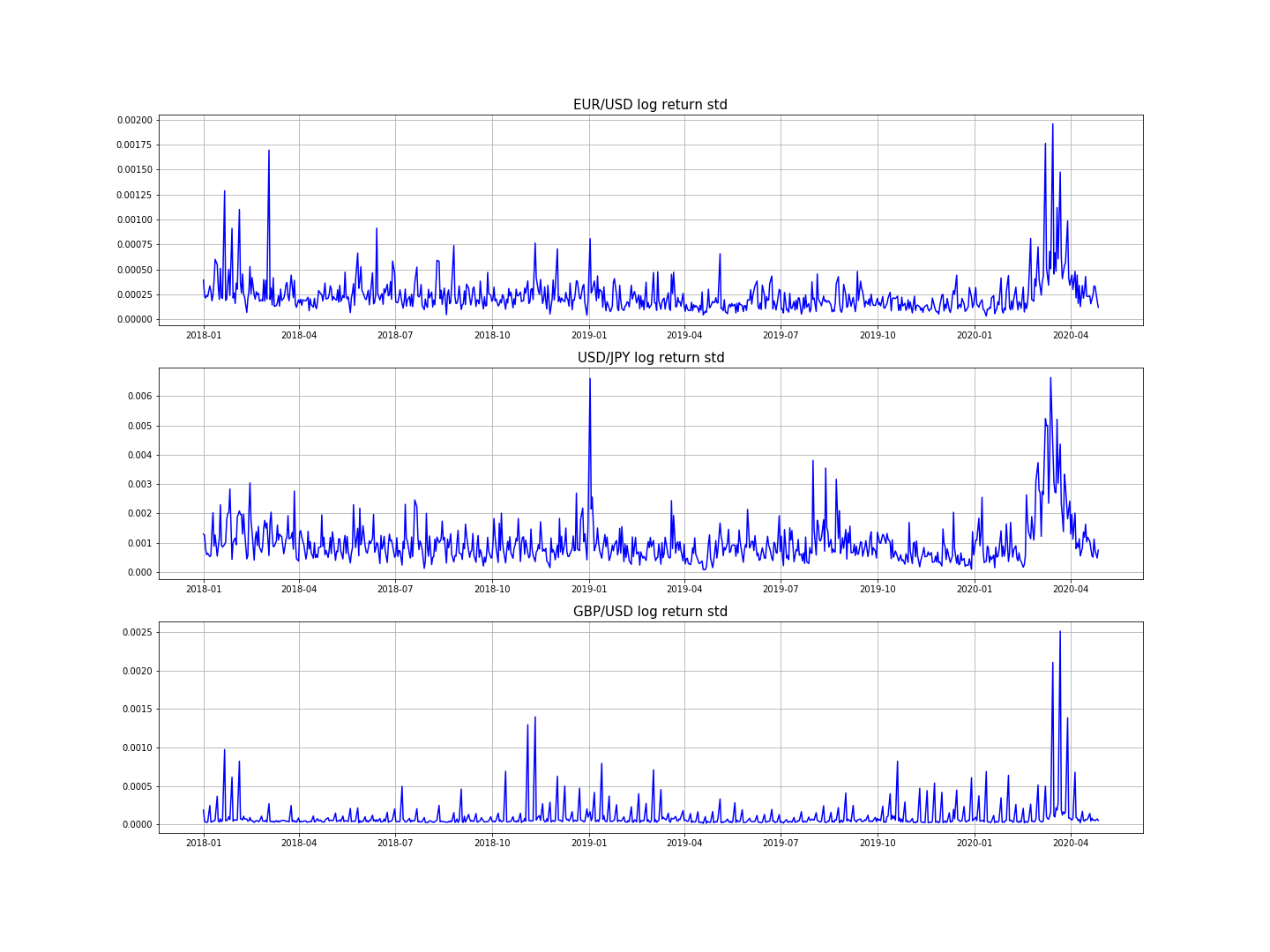}
    \caption{Daily volatility of currency pairs' log return}
    \label{ts_1}
\end{figure}

Currency pairs taken into consideration in this paper are freely floating and very liquid EUR/USD, GBP/USD and USD/JPY (figure \ref{ts_1}). For each of them we have statistical information for the last 14380 hours (from the the beginning of 2018 to 27-th April 2020), which includes open, high, low and close prices. \\ 

\begin{figure}
    \centering
    \includegraphics[scale = 0.4]{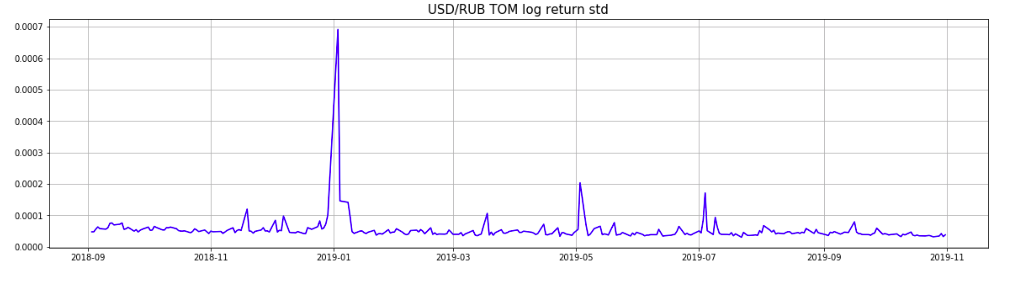}
    \caption{Daily volatility of USD/RUB TOM log return}
    \label{ts_rub}
\end{figure}

Moreover, we will analyze fairly unique tick data with prices for each trade in USD/RUB (TOM) instrument traded at foreign exchange market of Moscow Exchange. This data set includes more than 15 million observations and covers time period from September 2018 to October 2019, meaning that we can estimate daily volatility on roughly 58 000 observations (figure \ref{ts_rub}). \\

\addcontentsline{toc}{subsection}{2.2 Singular Spectrum Analysis}
\subsection*{\centering{\textbf{2.2 Singular Spectrum Analysis}}}

Now, when we are precise with the main hypothesis, task type (classification), data and metrics, we can move on to the first model - SSA.\\

\begin{figure}
    \centering
    \includegraphics[scale = 0.50]{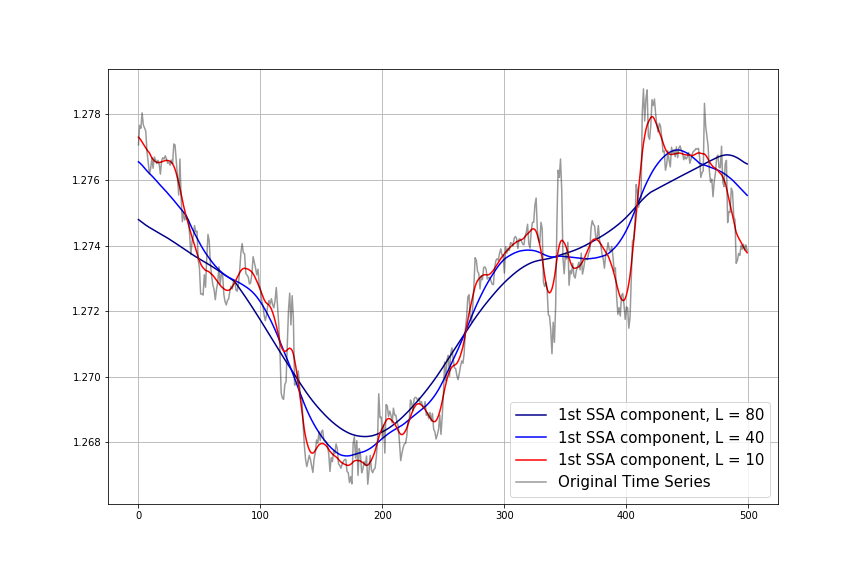}
    \caption{SSA time series decomposition with different window lengths}
    \label{ssa_1}
\end{figure}

The precision of time series approximation in singular spectrum analysis, which mechanics was described above, is heavily dependent on parameter $L$, which is usually called window length. Simply speaking, if $L$ is set to be big enough, one receives more components of initial time series, and each component carries a smaller proportion of initial information. When smaller $L$ is chosen by the researcher, each component presents a bigger proportion of information, and, hence tends to approximate original time series in more detail. Effects of different values of $L$ for financial series is presented on figure \ref{ssa_1}. As it is clearly seen, first component (which is the most valuable one in terms of information load) for $L = 10$ colored in red resembles original series precisely, while first SSA component for $L = 80$ reproduces only the direction without any intermediate deviations. \\

We will base SSA volatility modeling on the comparison of the dynamics of the $i-th$ SSA component $\mathbb{S}_i$ and original volatility process. Now, several technical questions regarding SSA computation arise: what proportion of data (moving window $\mathbb{W}$) should be used? What is the most proper SSA window length $L$? Moreover, we should determine the most becoming SSA component $\mathbb{S}_i$ for comparison. To pick the right combination for each currency pair $\mathbb{C} \in \{EUR/USD, GBP/USD, USD/JPY, USD/RUB\}$ we design SSA model with rolling window size $\mathbb{W} \in \{20, 30, 40, 50, 60, 70, 80, 90, 100, ..., 400\}$ and window length $L \in \{3, 4, 5, ..., 10\}$. After that, for each time step $t_i$ mean value of original volatility time series $T$ is compared with mean value of corresponding SSA trend $\mathbb{S}_i, i \in \{0, 1, 2\}$. The volatility forecast for time step $i+1$ is derived in the following way (equation \ref{ssa_forecast}):\\

\begin{equation} \label{ssa_forecast}
    \widehat{\sigma}_{i+1} = sign(\frac{1}{\mathbb{W}} \sum\limits_{t=i-\mathbb{W}}^{\mathbb{W}} \mathbb{S}_t - \frac{1}{\mathbb{W}} \sum\limits_{t=i-\mathbb{W}}^{\mathbb{W}} T_t)
\end{equation}

Thus, if for the current rolling window observed mean return volatility is lower than mean value of corresponding SSA trend, we expect future volatility levels to increase. Witnessed classification results after iterating through all currency pairs and SSA parameter combinations are shown in table \ref{ssa_res}. \\

\begin{center}
\begin{table}
    \begin{center}
    \caption{SSA forecasting results}
    \label{ssa_res}
    \vspace{5mm} 
    \begin{tabular}{|c|c|c|c|c|c|}
        \hline
        $\mathbb{C}$ & $L$ & $S$ & $\mathbb{W}$ & Accuracy score & F1 score\\ 
        \hline
        EUR/USD & 3 & 0 & 90 & 0.64 & 0.65 \\ 
        \hline
        GBP/USD & 3 & 0 & 20 & 0.66 & 0.69 \\ 
        \hline
        USD/JPY & 3 & 0 & 30 & 0.55 & 0.54 \\
        \hline
        USD/RUB & 7 & 0 & 40 & 0.62 & 0.63 \\
        \hline
    \end{tabular}
    \end{center}
\end{table}
\end{center}

\addcontentsline{toc}{subsection}{2.3 Dynamical System Stability Analysis}
\subsection*{\centering{\textbf{2.3 Dynamical System Stability Analysis}}}

As we have discussed earlier, the concept of a dynamical system's trajectory may serve as a good tool for financial modeling. One way to better understand the nature and relevant properties of a dynamical system is to estimate Lyapunov exponents, which act as a reliable measure of instability \citep{wolf1985determining}. This parameter indicates the speed of divergence of two paths with infinitesimally different initial conditions as the time passes. The average divergence at time $t$ is presented in \citet{rosenstein1993practical} (equation \ref{aver_divergence})

\begin{equation} \label{aver_divergence}
    d(t) = Ce^{\lambda_1t}
\end{equation}
where $C$ is a constant used for normalization and $\lambda_1$ is the biggest Lyapunov exponent from the Lyapunov spectrum. On the one hand, if two initially close trajectories tend to diverge with exponential speed, then the system swiftly becomes chaotic, unstable and unpredictable (in this case $\lambda_1$ is positive). On the other hand, when $\lambda_1$ is negative, trajectory is asymptotically stable (two close paths converge) and its behavior is thought to be predictable \citep{wolf1985determining}.\\

If we explicitly know the system of equations depicting the dynamical system, there is no problem in computing the whole Lyapunov spectrum, as precise techniques are discussed in \citet{malinetskii2000modern} and \citet{strogatz2001nonlinear}. However, when the only thing known for sure is the experimental time series (undoubtedly, there is some external noise effecting experimental data, which involves the analysis and interpretation of results), the task of finding even the largest Lyapunov exponent from the spectrum becomes challenging. A number of physicists have addressed this problem and derived different algorithms which differ a lot in terms of results acquired. Two good examples of robust technique are papers by \citet{rosenstein1993practical} and \citet{eckmann1986liapunov}. We will use the former algorithm as it is specifically designed for small data sets (with thousands of observations) and has yielded some promising results for dynamical systems studied in the paper. Additionally, \citet{rosenstein1993practical} approach is implemented in pyhton library $nolds$ by Christopher Schölzel for the tasks of bio-informatics.\\

\begin{figure}
    \centering
    \includegraphics[scale = 0.50]{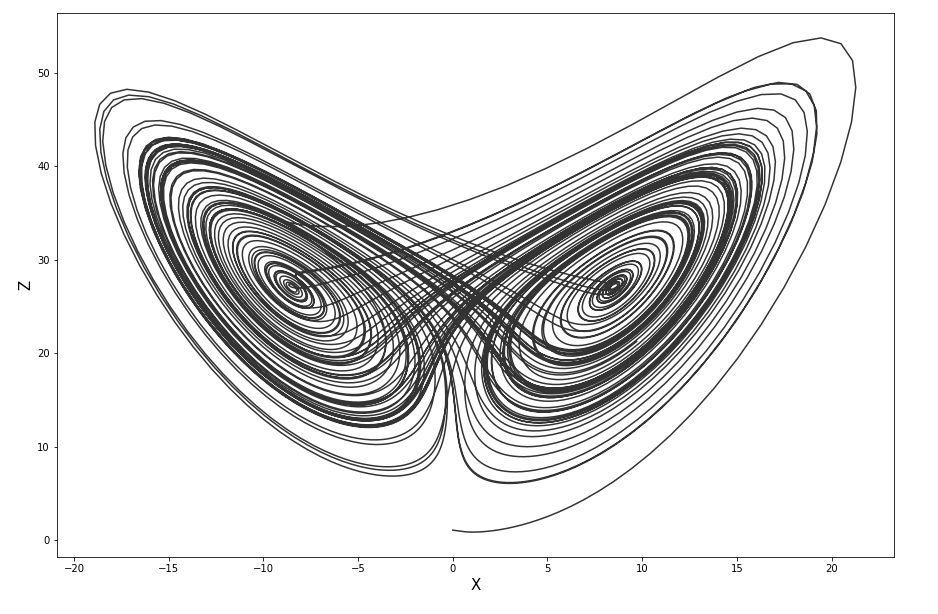}
    \caption{Lorenz Attractor in 2D space}
    \label{lorenz_attractor}
\end{figure}

First, we should determine the minimum length of experimental time series for which the first Lyapunov exponent is close to its fundamental value. As the size of time series grows, the approximation of $\lambda$ becomes more accurate, but we can stop at some point, where this approximation stops changing significantly with the increase of amount of data used. For this purpose we will use classical dynamical system presented by \citet{lorenz1963deterministic} (system of equations \ref{lorenz_system}), which attractor projection on $x-z$ axes is depicted in figure \ref{lorenz_attractor}.\\

\begin{equation} \label{lorenz_system}
    \systeme*{\frac{dx}{dt} = \sigma(y-x),
    \frac{dy}{dt} = x(p-z) -y, 
    \frac{dz}{dt} = xy - \beta z}
\end{equation}

\begin{figure}
    \centering
    \includegraphics[scale = 0.50]{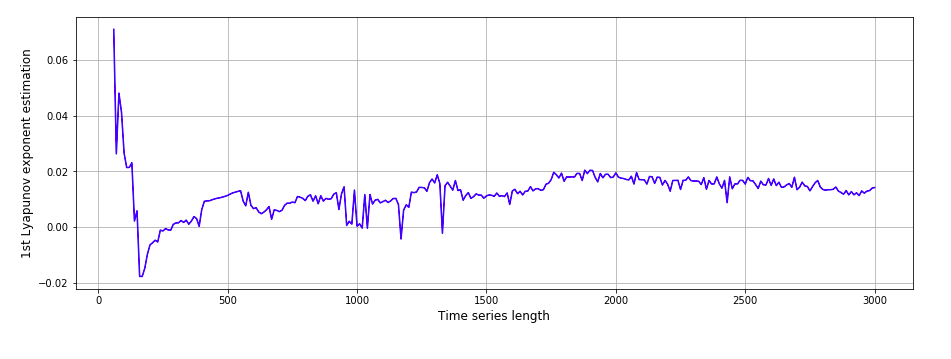}
    \caption{Estimation of $\lambda$ for Lorenz System}
    \label{lorenz_labbda}
\end{figure}

We then calculate the first Lyapunov exponent for $x$-axis time series using first $n \in \{60, 70, 80, ..., 3000\}$ points. The estimation of $\lambda$ seems to stabilize by 1500 data points used, as it is seen in figure \ref{lorenz_labbda}. \\

\begin{figure}%
    \centering
    \subfloat[1st differences of minute volatility distribution]{{\includegraphics[width=5cm]{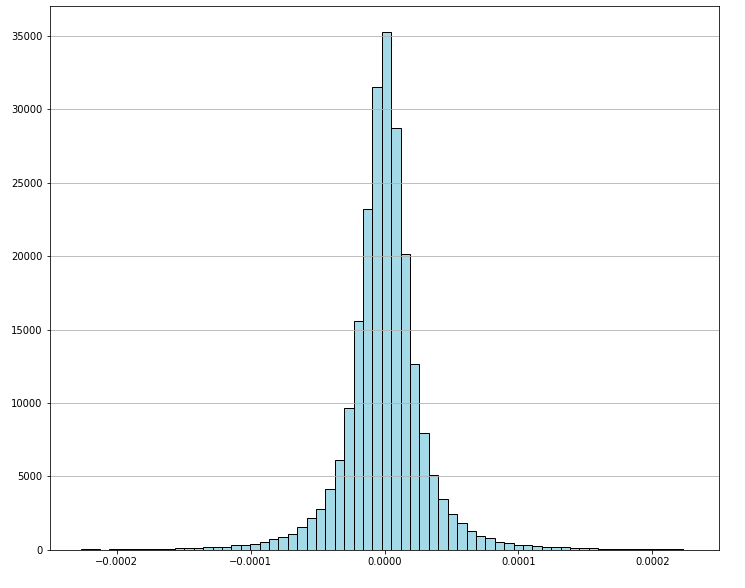} }}%
    \qquad
    \subfloat[minute volatiltiy distribution]{{\includegraphics[width=5cm]{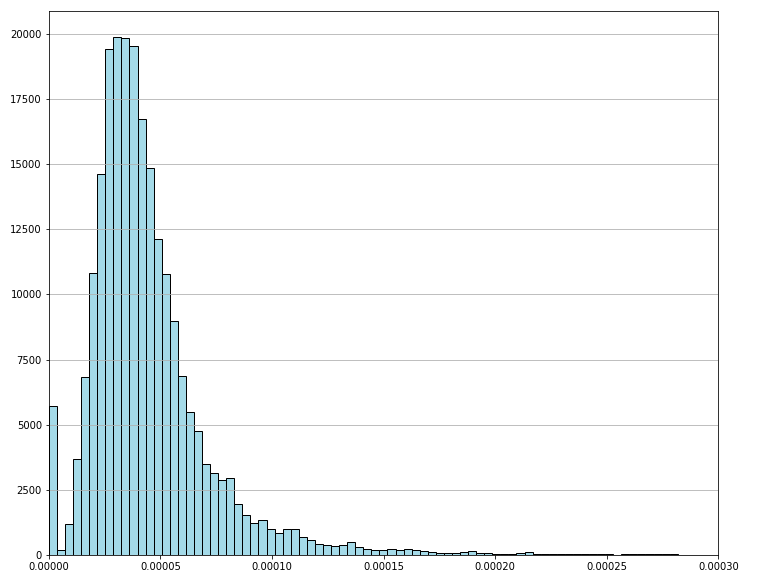} }}%
    \caption{Minute volatility}%
    \label{rub_vol_distr}%
\end{figure}

Thus, for the modeling purposes, we should set the moving window size of 1500 observations, which is impossible to do with EUR/USD, USD/JPY and GBP/USD as we posses only 723 daily volatility estimations. To solve this problem, we use USD/RUB TOM tick data for the estimation of observed volatility for each minute of a trading day with the formula \ref{volatility}. By doing this, we collect slightly less than 230 000 observations, which should be enough. The distribution of these volatility levels resembles log-normal distribution, although the left tail of histogram \ref{rub_vol_distr}b is heavy. When we look at first differences of minute volatility observations, they seem to be distributed as a Gaussian random variable (figure \ref{rub_vol_distr}a).\\ 

\begin{figure}
    \centering
    \includegraphics[scale = 0.50]{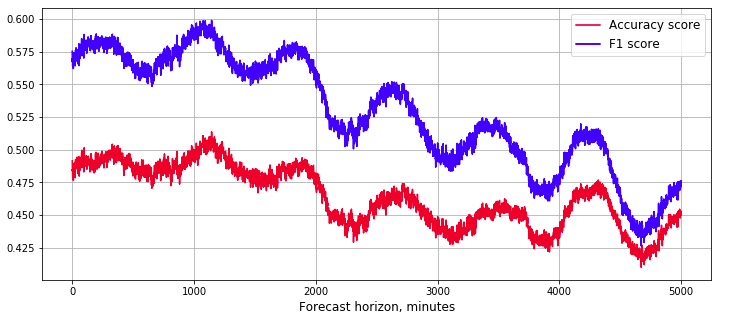}
    \caption{Accuracy and F1 scores for different forecast horizons}
    \label{interesting_behavior}
\end{figure}

For 10 000 points of observed volatility series we calculate $\lambda$ using \citet{rosenstein1993practical} method and assume that if $\lambda > 0$, then due to increasingly chaotic nature of dynamical system's attractor we can expect an increase in volatility over some period in the future. Alternatively, if $\lambda < 0$ and dynamical system can be described as stable for this mowing window, we look forward to seeing a decrease in future volatility. Still, we have one question to answer - what forecasting horizon should we use? To find an answer we construct forecasts for each time step of size $s \in \{1 min, 2 min, ..., 5000 min\}$. The acquired result (accuracy score and F1 score) is plotted in figure \ref{interesting_behavior}.\\

Surprisingly, accuracy and F1 metrics do not behave in a usual way, but rather move along some non-linear long-term trend with short-term 'seasonal' fluctuations. This may mean that although dynamical system's trajectory tends to evolve over time, there exists some periodicity with which trajectory gets closer to the initial condition. In other words, this 'spiral' development of volatility may enable us to find specific patterns in the process that one can find for different periods of time. Moreover, for the first 1500 minutes (which is exactly the moving window size) linear downside trend does not exist and both accuracy and F1 behave like a sine wave. This pretty unique outcome is definitely worth investigating in the further research on this topic.\\

Finally, we pick the most 'successful' horizon for the Lyapunov exponent based volatility forecasting. For 1144 minutes, which is approximately 19 hours, the results are significantly better than random choice in terms of F1 metric, but only slightly surpass random classifier in terms of accuracy score. The situation looks the same for 1 minute horizon, however, both metrics are slightly lower for 1 minute in comparison to 1144 minutes. The summary of Lyapunov exponent based modeling is presented in table \ref{lyapunov_res}.\\

\begin{center}
\begin{table}
    \begin{center}
    \caption{Lyapunov exponent forecasting results}
    \label{lyapunov_res}
    \vspace{5mm} 
    \begin{tabular}{|c|c|c|}
        \hline
        Forecast horizon& Accuracy score & F1 score\\ 
        \hline
        1 minute & 0.49 & 0.58 \\ 
        \hline
        1144 minutes & 0.51 & 0.60 \\ 
        \hline
    \end{tabular}
    \end{center}
\end{table}
\end{center}

\addcontentsline{toc}{subsection}{2.4 Generalized Autoregressive Conditional Heteroskedasticity}
\subsection*{\centering{\textbf{2.4 Generalized Autoregressive Conditional Heteroskedasticity}}}

Now, when we see the performance of singular spectrum analysis and dynamical system analysis in volatiltity modeling, we can move on to the baseline model of generalized conditional heteroskedasticity. For USD/JPY, GBP/USD and EUR/USD we will use daily observations, but for USD/RUB TOM we are forced to switch to the more frequent data (minute observations received from tick data) due to the small number of days (one year and one month) in the tick data set. GARCH is a one step forward model by definition, as it is seen from equation \ref{GARCH}, hence, for daily observations we will just estimate GARCH coefficients on a moving window (of size equal to the size of SSA moving window) and make a forecast for the next day. Concerning USD/RUB minute data, the procedure is a little bit more complicated. Here we will use bigger rolling window of size 1500 as it was done for the Lyapunov exponents and then produce two forecasts: one step (minute) forward forecast and 1144 steps forward forecast, which is based on forecasts for the previous minutes. This is done to have an opportunity to compare GARCH with both SSA and dynamical systems.\\

\begin{figure}
    \centering
    \includegraphics[scale = 0.4]{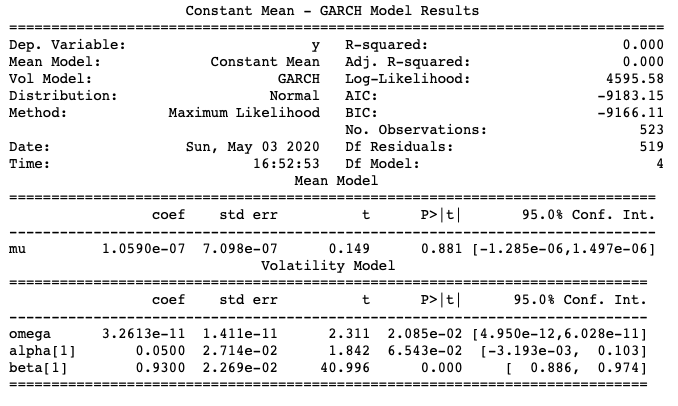}
    \caption{GARCH specification for EUR/USD train data}
    \label{garch_eur_spec}
\end{figure}

For three currency pairs with daily volatilities, initial time series of logarithmic returns is split into train (first 523 points) and test (residual 200 points) sets. Before making any forecasts we investigate overall suitability of GARCH for log return series. Classical approach of quantitative finance states that auto-regressive moving average model is a necessary preliminary step before GARCH, but in our case all ARMA models with adequate number of lags have statistically insignificant coefficients, which forces us to subtract mean logarithmic return from time series, which is already very close to zero, and pass it as an input to GARCH. For all three currency pairs there is only a modest improvement in information criteria with the growth of GARCH lags, so we decide to focus on a popular GARCH(1,1) model \citep{terasvirta2009introduction}. Models' specifications for train sub sample are given in the appendix (figures \ref{garch_jpy_spec}, \ref{garch_gbp_spec}, \ref{garch_eur_spec}). GARCH residuals analysis is also  presented in detail by figures \ref{garch_eur_resid}, \ref{garch_gbp_resid}, \ref{garch_jpy_resid} in appendix. As one can see, autoregressive conditional heteroskedasticity model with $p=q=1$ has failed to find all the dependencies in the log return data, as residuals of all three models for train data have significant autocorrelation. Surprisingly, we did not manage to get rid of the autocorrelation even when higher order lags $p$ and $q$ were included. Table \ref{garch_res} shows results acquired for moving window experiment. While accuracy scores are worse than any attempt with SSA or Lyapunov exponents, F1 score reveals bad performance of GARCH in regard to precision and recall. It is highly important to mention that although random binary guessing also yields $F1 = 0.5$ as it is clearly seen from equation \ref{f1}, simple forecast reversion of GARCH forecasts (from 1 to 0 and from 0 to 1) will not result in $F1$ score of $1 - F1_{old}$ as it happens with accuracy score. Thus, results for three currency pairs with daily observations indicate unconventional methods' dominance over GARCH.

\begin{center}
\begin{table}
    \begin{center}
    \caption{GARCH daily forecasting results}
    \label{garch_res}
    \vspace{5mm} 
    \begin{tabular}{|c|c|c|}
        \hline
        Currency pair & Accuracy score & F1 score\\ 
        \hline
        EUR/USD & 0.39 & 0.05 \\ 
        \hline
        GBP/USD & 0.37 & 0.03 \\ 
        \hline
        USD/JPY & 0.50 & 0.17 \\ 
        \hline
    \end{tabular}
    \end{center}
\end{table}
\end{center}

\begin{figure}%
    \centering
    \subfloat[ACF]{{\includegraphics[width=5cm]{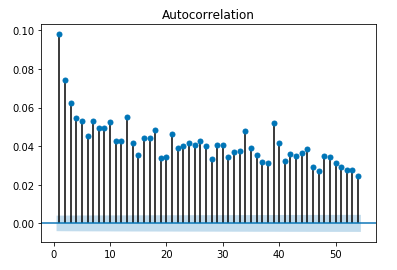} }}%
    \qquad
    \subfloat[PACF]{{\includegraphics[width=5cm]{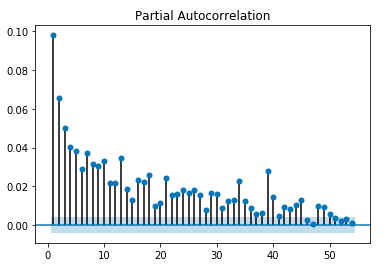} }}%
    \caption{USD/RUB TOM autocorrelation functions}%
    \label{rub_acf_pacf}%
\end{figure}

Autocorrelation and partial autocorrelation functions for USD/RUB TOM logarithmic returns with 1 minute time step indicate statistically significant dependency of observations up to the 50-th lag (figure \ref{rub_acf_pacf} a, b). Alike other currency pairs, we move from absolute values of log returns to residuals of ARMA(0, 4) model, which is the most suitable one from the set of ARMA models with $p$ and $q$ up to 5 in terms of AIC and BIC information criteria. After that, as it was described earlier, residuals are passed to GARCH(1, 1). This time we can use bigger training subset of size 1500, which would probably enhance quality of forecasts. Going through the data with larger moving window we, firstly, find the most becoming GARCH weights and, secondly, make one step forward forecast and 1144 steps forward forecast, which is simply 1144 small forecasts, each based on the previous ones as an input variables. The results depicted in table \ref{garch_res_rub} show that GARCH quality has indeed improved compared to the experiment with other currencies. However, both metrics make it clear that GARCH still underperforms and can not reach the level of Lyapunov exponent method, to the larger extent considering F1 metric. \\

\begin{center}
\begin{table}
    \begin{center}
    \caption{GARCH minute forecasting results}
    \label{garch_res_rub}
    \vspace{5mm} 
    \begin{tabular}{|c|c|c|}
        \hline
        Forecast horizon & Accuracy score & F1 score\\ 
        \hline
        1 minute & 0.48 & 0.31 \\ 
        \hline
        1144 minutes & 0.47 & 0.46 \\ 
        \hline
    \end{tabular}
    \end{center}
\end{table}
\end{center}

Thus, in both tests with different currency pairs and different data frequencies alternative methods of volatility modeling have proved to be more reliable than classical GARCH, at least in the task of binary classification. Recorded accuracy and F1 scores, which unambiguously compare approaches, present no reasons to reject $\mathbb{H}_0$ hypothesis. There are now more reasons to believe that iconoclastic methods like singular spectrum analysis and estimation of dynamical system's stability, which are widely used in physics, biology and other natural sciences can and should be implemented in financial econometrics, as they are highly probable to be useful for financial risk management, portfolio composition and many other tasks.

\newpage
\addcontentsline{toc}{section}{Conclusions}
\section*{\centering{\textbf{Conclusions}}}

In this paper we have investigated different methods of volatility modeling and forecasting. Classical GARCH model, which has been widespread in the econometric discourse since the second half of 1980s, was compared with understudied by economists singular spectrum analysis and dynamical system's stability analysis, which proved to be useful in natural sciences and could have performed better than GARCH according to our $\mathbb{H}_0$ hypothesis. Mentioned approaches were tested on a binary classification task, where class $I$ denoted an increase in observed volatility for the next time step and class $II$ stood for the decrease in volatility level. For the precise computation of models' performance accuracy and F1 metrics were used. Our data set included four currency pairs: EUR/USD, USD/JPY, GBP/USD, for which daily volatility observations were available, and unique high frequency USD/RUB TOM tick data provided by Moscow Exchange. For the latter, minute volatility observations were constructed.\\

For singular spectrum analysis, which is a technique splitting initial time series into trends by trajectory matrix decomposition, we have contrasted dynamics of the first trend, being the most informative one, to the dynamics of initial volatility time series on the most becoming moving window, which length was found by a simple search among all possible variants. If the mean value of reconstructed SSA trend was greater than mean value of original time series over specific period of time (equal to the moving window length), we have expected to see growth in observed volatility and vice-versa. \\

For the next experiment, we have assumed that observed volatility is a representation of the latent complex dynamical system, which can be described with the system of differential equations. This dynamical system has a number of paths or trajectories, which in turn could be stable or unstable. The more these paths are unstable, the more chaotic the system is, which makes the growth of the overall volatility more probable. Determination of the trajectory stability state is accomplished by the analysis of the Lyapunov spectrum that can be easily found analytically if one knows the system of differential equations explicitly. In our case, when only 'experimental' time series data is visible, Lyapunov exponents are computed using available information with the help of different methods proposed by scientists. The verdict regarding system stability was made by the analysis of the approximated first Lyapunov exponent. If the system was unstable, we have assumed growth in volatility in the future (up to 1144 steps forward), but if the system was found to be stable, an opposite process was anticipated.\\

Finally, GARCH model was used for the volatility modeling of all currency pairs mentioned above. As GARCH is a one step forward model by definition, there were no difficulties making a one step forward forecast, but we were forced to construct a 1144-step volatility prediction to compare GARCH with dynamical system analysis, using forecasts for previous time steps as an input.\\

Experiments have shown that SSA and dynamical system stability analysis indeed perform better than orthodox GARCH, at least for studied classification task specification and data used. While accuracy score of GARCH was significantly worse than random classifier for USD/JPY, GBP/USD, EUR/USD and slightly worse for USD/RUB, F1 score fully revealed its weakness in comparison to SSA and dynamical systems, which outperformed both GARCH and random classifier. Thus, we can not reject $\mathbb{H}_0$ hypothesis and proposed methods are likely to be useful for the problems of financial risk management and overall FX volatility modeling.\\

Without a doubt, this paper has its drawbacks and weak points that can be improved in further research. Firstly, SSA performance can be enhanced by the preliminary use of independent component analysis, which was discussed earlier in this paper. Secondly, we could have used break point analysis for more intelligent determination of moving window length. This technique is capable of finding break points in time series when the distribution of observations changes. This can potentially make classification quality of SSA and dynamical systems better. Finally, direct comparison of SSA and dynamical system's stability analysis was outside the scope of this research and seems to be a reasonable direction of further study.\\

\newpage

\addcontentsline{toc}{section}{References}
\renewcommand\refname{}
\section*{\centering{\textbf{References}}}
\bibliography{references}

\bibliographystyle{apalike}

\newpage
\addcontentsline{toc}{section}{Appendix}
\section*{\centering{\textbf{Appendix}}}

\begin{figure}[!htb]
    \centering
    \includegraphics[scale = 0.4]{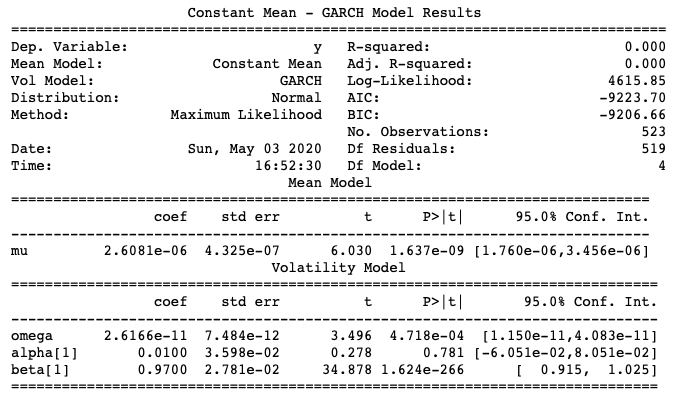}
    \caption{GARCH specification for USD/JPY train data}
    \label{garch_jpy_spec}
\end{figure}

\begin{figure}[!htb]
    \centering
    \includegraphics[scale = 0.50]{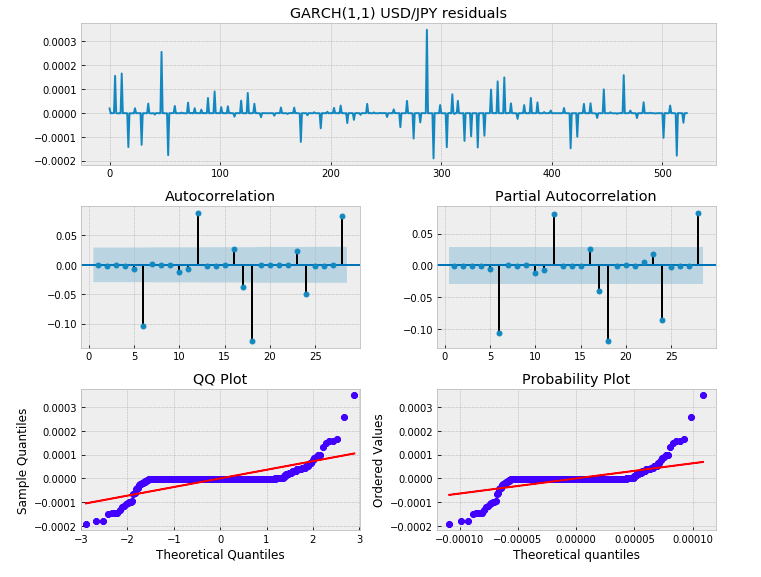}
    \caption{GARCH residuals for USD/JPY train data}
    \label{garch_jpy_resid}
\end{figure}

\begin{figure}
    \centering
    \includegraphics[scale = 0.4]{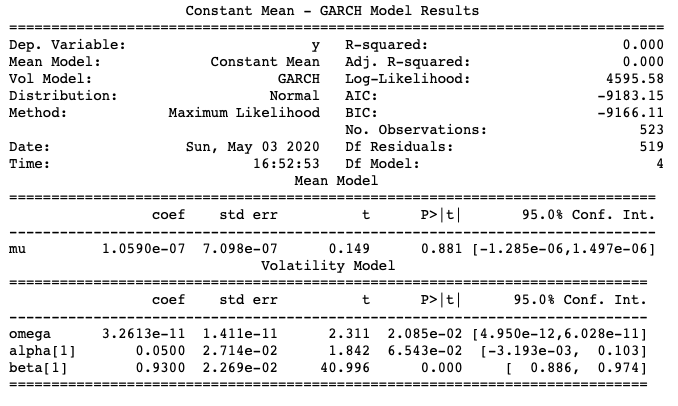}
    \caption{GARCH specification for GBP/USD train data}
    \label{garch_gbp_spec}
\end{figure}

\begin{figure}
    \centering
    \includegraphics[scale = 0.50]{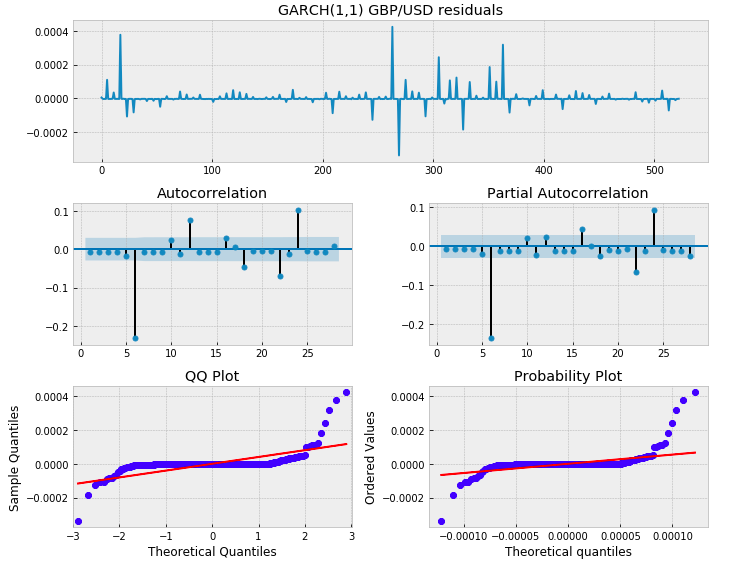}
    \caption{GARCH residuals for GBP/USD train data}
    \label{garch_gbp_resid}
\end{figure}

\begin{figure}
    \centering
    \includegraphics[scale = 0.50]{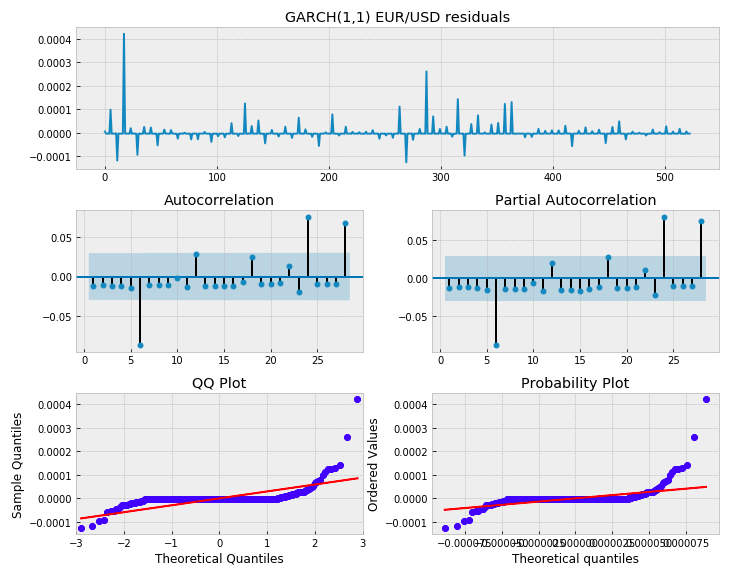}
    \caption{GARCH residuals for EUR/USD train data}
    \label{garch_eur_resid}
\end{figure}

\end{spacing}

\end{document}